\documentstyle[12pt]{article}

\newtheorem{theorem}{Theorem}
\newtheorem{lemma}{Lemma}
\newtheorem{cor}{Corollary}
\newcommand{\proof}{\noindent{\bf Proof.}\ }
\begin{document}

\newcommand{\lap}{\bigtriangleup}
\def\be{\begin{equation}}
\def\ee{\end{equation}}
\def\bea{\begin{eqnarray}}
\def\eea{\end{eqnarray}}
\def\beas{\begin{eqnarray*}}
\def\eeas{\end{eqnarray*}}
\def\n#1{\vert #1 \vert}
\def\nn#1{{\Vert #1 \Vert}}
 
\def\R{{\rm I\kern-.1567em R}}
\def\N{{\rm I\kern-.1567em N}}
 
\def\supp{\mbox{\rm supp}\,}
\def\dist{\mbox{\rm dist}\,}

\def\ekin{E_{\rm kin}}
\def\epot{E_{\rm pot}}

\def\C{{\cal C}}
\def\F{{\cal F}}
\def\G{{\cal G}}
\def\H{{\cal H}}
\def\Hc{{\cal H_C}}
\def\Hcr{{{\cal H}_{\cal C}^r}}
\def\P{{\cal P}}
\def\V{{\cal V}}
\def\r{\rho}

\def\prfe{\hspace*{\fill} $\Box$

\smallskip \noindent}

\title{Reduction and a concentration-compactness principle
       for energy-Casimir functionals}

\author{ Gerhard Rein\\
         Mathematisches Institut 
         der Universit\"at M\"unchen\\
         Theresienstr. 39\\
         80333 M\"unchen, Germany}
\date{}
\maketitle

\begin{abstract}
Energy-Casimir functionals are a useful tool for the 
construction of steady states and the analysis of their nonlinear
stability properties for a variety of conservative systems in 
mathematical physics. Recently, Y.~Guo and the author 
employed them to construct stable steady states
for the Vlasov-Poisson system in stellar dynamics,
where the energy-Casimir functionals act on number density 
functions on phase space.
In the present paper we construct natural, reduced functionals
which act on mass densities on space and study compactness
properties and the existence of minimizers in this context.
This puts the techniques developed by Y.~Guo and the author
into a more general framework. We recover the 
concentration-compactness
principle due to P.~L.~Lions \cite{L} in a more specific setting
and connect our stability analysis with the one of 
G.~Wolansky \cite{Wo}.

\end{abstract}

\section{Introduction}
\setcounter{equation}{0}

The purpose of the present paper is to investigate the
compactness properties and existence of minimizers of
certain functionals which appear naturally in the stability
analysis of various systems in kinetic theory.
Given a large ensemble of particles which interact by
gravitational attraction
we consider energy-Casimir functionals
which are defined on the space of phase space density functions,
and certain reduced versions of these which are defined
on the space of spatial density functions. This reduction 
procedure should put the techniques developed in 
\cite{G1,G2,GR1,GR2,GR3,R1,R2}
into a more general framework and make them applicable
to problems outside kinetic theory. However, to be specific
we start by recalling the Vlasov-Poisson system which describes
the time evolution of a large ensemble of particles interacting
by the gravitational field which they create collectively:   
\[
\partial_t f + v \cdot \partial_x f - \partial_x U \cdot 
\partial_v f = 0 ,
\]
\[ 
\lap U = 4 \pi\, \r,\ \lim_{|x| \to \infty} U(t,x) = 0 , 
\]
\[
\r(t,x)= \int f(t,x,v)dv.
\]
Here the dynamic variable is the number density $f=f(t,x,v)$ of the 
ensemble in phase space, $x, v \in \R^3$ denote position and 
velocity, $\r = \r(t,x)$ is the spatial mass density induced by
$f$, and $U=U(t,x)$ is the induced gravitational potential.
It is straight forward to check that
\[
\int\!\!\int Q(f(t,x,v))\,dv\,dx 
+ \frac{1}{2} \int\!\!\int |v|^2 f(t,x,v)\,dv\,dx
- \frac{1}{2} \int\!\!\int \frac{\r(t,x)\, \r(t,y)}{\n{x-y}}
\, dx\, dy
\]
is conserved along solutions for any suitable scalar function $Q$.
The first part, which is conserved by itself, is a so-called Casimir 
functional, the second is the kinetic and the third part
the potential energy of the system. When viewed as a functional on
phase space densities $f=f(x,v)\geq 0$ we denote this functional by
$\Hc$. It is fairly straight forward to see that
any minimizer of this functional subject to the constraint
\[
\int\!\!\int f(x,v)\,dv\,dx = M
\]
with prescribed total mass $M>0$ is a steady state of the 
Vlasov-Poisson system. It is far less obvious that such minimizers 
exist and that they are nonlinearly stable. When analyzing the 
minimization problem
\be \label{vp}
\Hc(f_0)=\inf \left\{ \Hc (f) | 
f \geq 0, \int\!\!\int f \,dv\,dx = M \right\},
\ee
one needs to make sure that one can pass to the limit in
the (quadratic) potential energy along a minimizing sequence.
Obviously, the potential energy is not a functional
of $f$ itself, but of the induced spatial density
$\rho$, and the crucial question
is how along a minimizing sequence the spatial density can
or cannot split into parts or spread uniformly in space.
This was analyzed in the context of the Vlasov-Poisson
system and with various variations in 
\cite{G1,G2,GR1,GR2,GR3,R1,R2}.

In the present paper we want to bring out the basic 
mechanism more clearly and in a framework not restricted
to kinetic theory. To this end we construct in 
the next section
a reduced version $\Hcr$ of the energy-Casimir functional 
$\Hc$, which will be defined on spatial densities $\r$,
\[
\Hcr (\r) = \int \Phi (\r(x))\, dx
- \frac{1}{2} \int\!\!\int \frac{\r(x)\, \r(y)}{\n{x-y}}
\, dx\, dy
\] 
with $\Phi$ a function determined by $Q$, which is convex if $Q$
is convex.
Then we explore the relation between
the variational problems    
\be \label{vpr}
\Hcr(\r_0)=\inf \left\{ \Hcr (\r) | 
\rho \geq 0, \int \rho \,dx = M \right\}
\ee
and (\ref{vp}), in particular we will show how a minimizer
of the reduced problem (\ref{vpr}) induces a minimizer
of (\ref{vp}).
In the third section we reformulate the techniques developed for (\ref{vp})
in the framework of (\ref{vpr}) and obtain the existence
of a minimizer $\r_0$ under appropriate conditions on $\Phi$. 
In particular, we prove
the essential part of the concentration-compactness principle
due to P.-L.~Lions \cite{L} by a more direct method based on
scaling and splitting.
In the last section we discuss the role of symmetries in
the problem and point out some applications and extensions
of our results.
An example of a function $\Phi$ which satisfies
all the necessary assumptions is $\Phi(\r) = \r^{1+1/n}$
with $0<n<3$. In this case the potential $U_0$ induced by 
a minimizer $\r_0$ 
is a solution of the semilinear elliptic problem
\[
\lap U_0 = (E_0-U_0)_+^n,\ 
\lim_{\n{x}\to \infty } U_0(x)=0
\]
where $(\cdot)_+$ denotes the positive part and $E_0$ is some constant . 
This equation is sometimes referred
to as the Emden-Fowler equation and appears naturally in the study of
self-gravitating fluid balls. Throughout this paper we restrict ourselves
to the case of space dimension 3; extending these techniques to other
space dimensions by adjusting various exponents is easy.

\section{Reduction of energy-Casimir functionals}
\setcounter{equation}{0}

For a measurable function $f=f(x,v)$ we define
\[
\rho_f (x):= \int f(x,v)\, dv,\ x \in \R^3,
\]
and
\[  
U_f :=  - \rho_f \ast \frac{1}{|\cdot|}.
\]
Next we define
\beas
\ekin (f)
&:=&
\frac{1}{2} \int\!\!\int |v|^2 f(x,v)\,dv\,dx \\
\epot (f)
&:=&
- \frac{1}{8\pi} \int |\nabla U_f (x)|^2 dx = 
- \frac{1}{2} \int\!\!\int\frac{\rho_f(x) \rho_f(y)}{|x-y|}dx\,dy,
\eeas
---$\epot$ can equally well be viewed as a functional of $\r$
instead of $f$---, and
\[
\Hc(f) :=
\C(f) + \ekin(f)  + \epot (f), 
\]
where
\[
\C(f)
:=
\int\!\!\int Q(f(x,v))\,dv\,dx
\]
and $Q$ is a given function satisfying the following

\smallskip
\noindent {\bf Assumption on $Q$}:\\ 
$Q \in C^1 ([0,\infty[)$ is strictly convex, $Q(0)=Q'(0)=0$,
and $Q(f)/f \to \infty,\ f \to \infty$.

\smallskip\noindent
In particular, this implies that $Q\geq 0$ and 
$Q':[0,\infty[ \to [0,\infty[$ is one-to-one and onto.

We study the following variational problem: Minimize $\Hc$
over the set
\be \label{spacedef}
\F_M := \Bigl\{ f \in L^1_+ (\R^6) 
\mid
\C(f) + \ekin(f) < \infty,\ \r_f \in L^{6/5}(\R^3),
\ \int\!\!\int f = M \Bigr\},
\ee
where $M>0$ is prescribed and $L^1_+ (\R^6)$  denotes the set of
a.~e.\ nonnegative functions in $L^1 (\R^6)$.
Note that since $\r_f \in L^{6/5}(\R^3)$
the convolution defining $U_f$ exists in $L^6(\R^3)$
with $\nabla U_f \in L^2(\R^3)$ 
according to the extended Young's inequality, and the
potential energy of $\r_f$ is finite.

In order to guarantee the existence of a minimizer
we will require additional growth conditions on $Q$ to be introduced
later; at the moment $\epot(f)$ could be minus infinity for $f \in \F_M$.  
A typical example of a function for which there exists a minimizer is 
$Q(f)=f^{1+1/k}$ with $0<k<3/2$.   

To obtain a reformulation in terms of spatial densities $\r$
which captures the essential properties of this 
variational problem we proceed as follows.
For $r \geq 0$ we define
\be \label{vdef}
\G_r :=\left\{g \in L^1_+ (\R^3) | 
\int\left(\frac{1}{2}|v|^2 g (v) + Q(g(v))\right)\,dv < \infty,\
\int g(v)\, dv = r \right\} 
\ee
and
\be \label{phidef}
\Phi(r):=\inf_{g \in\G_r} 
\int\left(\frac{1}{2}|v|^2 g (v) + Q(g(v))\right)\,dv.
\ee
In addition to the variational problem of minimizing
$\Hc$ over the set $\F_M$ we consider the
problem of minimizing the functional
\be \label{hcrdef}
\Hcr(\r)
:=
\int \Phi (\r(x))\, dx + \epot (\r)
\ee
over the set
\be \label{frdef}
\F_M^r
:=
\left \{\r \in L^{6/5} \cap L^1_+(\R^3) \mid \int \Phi (\r(x))\, dx < \infty ,
\ \int \r (x)\, dx = M
 \right\}
\ee
The relation between the minimizers of $\Hc$ and $\Hcr$ is the main
theme of this section, and
a remark on how we passed from $\Hc$ to $\Hcr$ 
can be found at the end of the section. 

\begin{theorem} \label{reduce} 
\begin{itemize}
\item[{\rm (a)}]
For every function $f \in \F_M$,
\[
\Hc(f) \geq \Hcr(\r_f),
\]
and if $f=f_0$ is a minimizer of $\Hc$ over $\F_M$ then equality holds.
\item[{\rm (b)}]
Let  $\r_0 \in \F_M^r$ be a minimizer of $\Hcr$ with
induced potential $U_0$.
Then there exists a Lagrange multiplier $E_0 \in \R$ such that
a.~e.,
\[
\r_0 = \left\{ 
\begin{array}{ccl}
(\Phi')^{-1}(E_0 - U_0)&,& U_0 < E_0 \\
0 &,& U_0 \geq E_0. 
\end{array}
\right.
\]
Denote by 
\[
E = E(x,v):=\frac{1}{2} |v|^2 + U_0(x)
\]
the energy of a particle at position $x$ with velocity $v$, and define
\[
f_0
:=
\left\{ 
\begin{array}{ccl}
(Q')^{-1}(E_0 - E) &,& E < E_0, \\
0 &,& E \geq E_0. 
\end{array}
\right.
\]  
Then $f_0 \in \F_M$ is a minimizer of $\Hc$.
\item[{\rm (c)}]
Now {\em assume} that $\Hcr$ has at least one minimizer in $\F_M^r$.
Then the following holds:
If $f_0\in \F_M$ is a minimizer of $\Hc$
then $\r_0 := \r_{f_0} \in \F_M^r$ is a minimizer of $\Hcr$,
this map is one-to-one and onto
between the sets of minimizers of $\Hc$
in $\F_M$ and of $\Hcr$ in $\F_M^r$ respectively,
and is the inverse of the map $\r_0 \mapsto f_0$
described in (b).  
\end{itemize}
\end{theorem}

\noindent
{\bf Remark.} This theorem does not exclude the possibility
that $\Hc$ has a minimizer but $\Hcr$ has none.
In the next section we show that under appropriate assumptions on $\Phi$
the reduced functional $\Hcr$ does have a minimizer,
and then the theorem guarantees that we recover all minimizers of
$\Hc$ in $\F_M$ by ``lifting'' the ones of $\Hcr$ as described in (b). 

Before we prove this theorem, we investigate the relation between
$Q$ and $\Phi$;
for a function 
$h:\R \to ]-\infty,\infty]$ we denote by
\[
h^\ast (\lambda):= \sup_{r \in \R} (\lambda \, r - h(r))
\]
its Legendre transform. Some of the results of the lemma below
will be relevant for the next section.

\begin{lemma} \label{phiprop}
Let $Q$ be as specified above, let $\Phi$ be defined by (\ref{vdef}),
(\ref{phidef}), and extend both functions by $+\infty$ 
to the interval $]-\infty,0[$. 
\begin{itemize}
\item[{\rm (a)}]
For $\lambda \in \R$,  
\[
\Phi^\ast (\lambda)
=
\int Q^\ast \left( \lambda - \frac{1}{2} |v|^2 \right)\, dv,
\]
and in particular,
$Q^\ast(\lambda)=0 = \Phi^\ast(\lambda)$ 
for $\lambda <0$.
\item[{\rm (b)}]
$\Phi \in C^1 ([0,\infty[)$ is strictly convex, and
$\Phi(0)=\Phi'(0)=0$.
\item[{\rm (c)}]
Let $k>0$ and $n=k+3/2$. As in the rest of the paper,
constants denoted by $C$ are positive, may depend on $Q$ or $M$,
and may change from line to line (or within one line).
\begin{itemize}
\item[{\rm (i)}] 
If $Q(f) = C\, f^{1+1/k},\ f \geq 0$, then 
$\Phi(\r) = C\, \r^{1+1/n},\ \r \geq 0$.
\item[{\rm (ii)}] 
If $Q(f)  \geq C\, f^{1+1/k},\ f \geq 0$ large, then 
$\Phi(\r) \geq C\, \r^{1+1/n},\ \r \geq 0$ large.
\item[{\rm (iii)}] 
If $Q(f)  \leq C\, f^{1+1/k},\ f \geq 0$ small, then 
$\Phi(\r) \leq C\, \r^{1+1/n},\ \r \geq 0$ small.
\end{itemize}
If the restriction to large respectively small values of $f$
can be dropped, then the corresponding restriction for $\r$
can be dropped as well.
\end{itemize}
\end{lemma} 

\proof
By definition,
\beas
\Phi^\ast (\lambda)
&=&
\sup_{r\geq 0} \left[ \lambda \, r - \inf_{g\in \G_r}
\int\left(\frac{1}{2}|v|^2 g(v) + Q(g(v))\right)\, dv\right]\\
&=&
\sup_{r\geq 0}
\sup_{g \in \G_r}
\int\left[\left(\lambda  -\frac{1}{2}|v|^2\right) g(v) 
- Q(g(v))\right]\, dv \\
&=&
\sup_{g\in L^1_+(\R^3)}
\int\left[\left(\lambda  -\frac{1}{2}|v|^2\right) g(v) 
- Q(g(v))\right]\, dv\\
&=&
\int\sup_{y\geq 0}
\left[\left(\lambda  -\frac{1}{2}|v|^2\right) y 
- Q(y)\right]\, dv =
\int Q^\ast\left(\lambda - \frac{1}{2}|v|^2\right)\, dv.
\eeas
As to the last-but-one equality, observe that both sides are obviously
zero for $\lambda \leq 0$. If $\lambda >0$ then for
any $g \in L^1_+(\R^3)$,
\[
\int\left[\left(\lambda  -\frac{1}{2}|v|^2\right) g(v) 
- Q(g(v))\right] \, dv
\leq
\int\sup_{y\geq 0}
\left[\left(\lambda  -\frac{1}{2}|v|^2\right) y 
- Q(y)\right]\, dv.
\]
If $\n{v} \geq \sqrt{2 \lambda}$ then 
$\sup_{y\geq 0} \left[ \cdots \right] = 0$, and 
for $\n{v} < \sqrt{2 \lambda}$ 
the supremum of the term in brackets is attained at 
$y=y_v:=(Q')^{-1}\left(\lambda  -\frac{1}{2}|v|^2\right)$.
Thus with
\[
g_0(v):= \left\{
\begin{array}{ccl}
y_v &,& \n{v} < \sqrt{2 \lambda},\\
0 &,& \n{v} \geq \sqrt{2 \lambda},
\end{array}
\right.
\]
we have
\beas
\int\sup_{y\geq 0}
\left[\left(\lambda  -\frac{1}{2}|v|^2\right) y 
- Q(y)\right]\, dv
&=&
\int\left[\left(\lambda  -\frac{1}{2}|v|^2\right) g_0(v) 
- Q(g_0(v))\right]\, dv\\ 
&\leq&
\sup_{g\in L^1_+(\R^3)}
\int\left[\left(\lambda  -\frac{1}{2}|v|^2\right) g(v) 
- Q(g(v))\right]\, dv,
\eeas
and part (a) is established.

Since $Q$ is strictly convex and lower semi-continuous as a function
on $\R$ with $\lim_{|f| \to \infty}Q(f)/|f| \to \infty$,
$Q^\ast \in C^1(\R)$, cf.\ \cite[Prop.~2.4]{MW}. 
Obviously, $Q^\ast (\lambda)=0$ for $\lambda \leq 0$,
in particular, $(Q^\ast)' (0)=0$. Also, $(Q^\ast)'$
is strictly increasing on $[0,\infty[$ since $Q'$ is strictly increasing
on $[0,\infty[$ with range $[0,\infty[$. Since for $|\lambda| < \lambda_0$
with $\lambda_0>0$ fixed the integral in the formula for
$\Phi^\ast$ extends over a compact set we may differentiate
under the integral sign to conclude that $\Phi^\ast \in C^1(\R)$ with
derivative strictly increasing on $[0,\infty[$.
This in turn implies the assertion of part (b).

Finally, $Q(f) \geq C\, f^{1+1/k},\ f \geq 0$ large, implies
that $Q(f) \geq C\, f^{1+1/k}-C',\ f \geq 0$. Thus 
\[
Q^\ast(\lambda)
\leq
C' + \sup_{f\geq 0} \left( f \lambda - C\, f^{1+1/k} \right)
=
C' + \frac{1}{1+k} \left(\frac{k}{C_Q\,(1+k)}\right)^k \lambda^{1+k}
,\ \lambda \geq 0,
\]
and
\beas
\Phi^\ast(\lambda)
&\leq&
C \lambda^{3/2} + C \int_{\n{v}\leq \sqrt{2 \lambda}}
\left( \lambda - \frac{1}{2} |v|^2\right)^{1+k} dv
=
C \lambda^{3/2} + C \int_0^\lambda E^{1+k} 
\sqrt{\lambda - E}\, dE \\
&=&
C' + C \, \lambda^{k+5/2} = C' + C \lambda^{1+n} ,\ \lambda \geq 0.
\eeas
This in turn yields the assertion on $\Phi$ in (ii).
The assertion in (i) is now obvious. As to (iii) note first that
for $\lambda \geq 0$ and small the corresponding supremum
is attained at small $f$'s, and thus
\[
Q^\ast (\lambda) \leq \sup_{f \geq 0} \left(\lambda \, f - C\, f^{1+1/k}\right)
= C \lambda^{1+k}.
\]
Thus still for $\lambda \geq 0$ small,
$\Phi^\ast (\lambda) \geq C \lambda^{1+n}$,
which in turn implies the assertion for $\Phi$.
\prfe   

We now prove the theorem above.

\noindent
{\bf Proof of Theorem~\ref{reduce}.}\\
{\em Proof of the inequality in part (a).} For $\r \in \F_M^r$
define
\be \label{frhodef}
\F_\r :=\{f \in \F_M |\r_f = \r \} .
\ee
Clearly, for $\r = \r_f$ with $f \in \F_M$,
\bea \label{intermest}
\C(f) + \ekin(f) 
&\geq& 
\inf_{\tilde f \in \F_\r}(\C(\tilde f) + \ekin(\tilde f)) \nonumber \\
&\geq&
\inf_{\tilde f \in\F_\r} 
\int \biggl[\inf_{g \in \G_{\r (x)}} 
\int \left(\frac{1}{2}|v|^2 g(v) + Q(g(v))\right)\,dv\biggr]\,dx \nonumber\\
&=&
\int \biggl[\inf_{g \in\G_{\r(x)}} 
\int \left(\frac{1}{2}|v|^2 g(v) + Q(g(v))\right)\,dv\biggr]\,dx \nonumber \\
&=&
\int \Phi(\r (x))\, dx,
\eea
and the inequality in part (a) is established. 

\noindent
{\em An intermediate assertion.} We claim that
if $f \in \F_M$ is such that up to sets of measure zero,
\be \label{euleq}
\left. 
\begin{array}{ccl} 
Q'(f) = E_0-E > 0 &,& 
\mbox{where}\ f>0 ,\\
E_0-E \leq 0 &,& \mbox{where}\ f = 0 .
\end{array}
\right\}
\ee
with $E$ defined as in (b) but with $U_f$ instead of $U_0$ 
and $E_0$ a constant then equality holds in part (a).

To prove this, observe that since $Q$ is convex, 
we have for a.~e.\ $x \in \R^3$
and every $g \in \G_{\r_f (x)}$,
\beas
\frac{1}{2}|v|^2 g(v) + Q(g(v)) 
&\geq&
\frac{1}{2}|v|^2 f(x,v) + Q(f(x,v))\\
&&
{} + \left(\frac{1}{2}|v|^2 + Q'(f(x,v))\right)\, (g(v) - f(x,v))\ 
\mbox{a.~e.}
\eeas
Now by (\ref{euleq}),
\beas 
&&\int\left(\frac{1}{2}|v|^2 + Q'(f)\right)\, (g - f)\, dv
=
\int_{\{f>0\}}\ldots + \int_{\{f=0\}}\ldots \\
&& \qquad \qquad
=
(E_0-U_f(x)) \int_{\{f>0\}} (g - f)\, dv 
+ \int_{\{f=0\}} \frac{1}{2} |v|^2 g\, dv\\
&& \qquad \qquad
= - (E_0-U_f(x)) \int_{\{f=0\}} (g - f)\, dv
+ \int_{\{f=0\}} \frac{1}{2} |v|^2 g\, dv\\
&& \qquad \qquad
=
\int_{\{f=0\}} (E - E_0)\,g\, dv \geq 0;
\eeas
observe that $g \geq 0$ and $\int g\,dv = \int f\, dv$
so $\int (g - f)\, dv = 0$.
Thus we see that
\beas
\Phi(\r_f(x))
&\geq&
\int\left(\frac{1}{2}|v|^2 f + Q(f)\right)\, dv\\
&\geq&
\inf_{g\in\G_{\r_f(x)}}
\int\left(\frac{1}{2}|v|^2 g + Q(g)\right)\, dv =
\Phi(\r_f(x))\ \mbox{a.~e.},
\eeas
and the proof of our intermediate assertion is complete. 

\noindent
{\em Proof of the equality assertion in (a).}
If $f_0\in \F_M$ is a minimizer of $\Hc$
then the Euler-Lagrange equation of the 
minimization problem implies that (\ref{euleq}) holds
for some Lagrange multiplier $E_0$; this can be proved as 
in \cite[Thm.~2]{GR3}.
Thus equality holds in (a) by the intermediate assertion,
and the proof of part (a) is complete.

\noindent
{\em Proof of part (b)}.
Let $\r_0 \in \F_M^r$ be a minimizer of $\Hcr$. Then the 
Euler-Lagrange equation yields the relation between
$\r_0$ and $U_0$. Let $f_0$ be defined as in (b). 
Then up to sets of measure zero,
\beas
\int f_0(x,v)\, dv 
&=&
\int_{\n{v} \leq \sqrt{2(E_0-U_0(x))}}
(Q')^{-1}\left(E_0-U_0(x)-\frac{1}{2}\n{v}^2\right)\, dv \\
&=&
(\Phi^\ast)'(E_0-U_0(x)) =
(\Phi')^{-1}(E_0-U_0(x)) = \r_0(x)
\eeas
where $U_0(x)< E_0$, and both sides are zero 
where $U_0(x)\geq E_0$. Thus $\r_0=\r_{f_0}$,
in particular, $f_0 \in \F_M$. By definition,
$f_0$ satisfies the Euler-Lagrange relation (\ref{euleq})
and thus by our intermediate assertion
$\Hc(f_0) = \Hcr(\r_0)$. Therefore again by part (a),
\[
\Hc(f) \geq \Hcr(\r_f) \geq \Hcr(\r_0) = \Hc (f_0),\ f \in \F_M,
\]
so that $f_0$ is a minimizer of $\Hc$, and the proof of part (b)
is complete.

\noindent
{\em Proof of part (c)}.
Assume that $\Hcr$ has a minimizer $\r_0 \in \F_M^r$ and define $f_0$
as above. Then part (a),
the fact that each $\r \in \F_M^r$ can 
be written as $\r =\r_f$
for some $f\in\F_M$, and our intermediate assertion imply that
\bea
\inf_{f \in \F_M} \Hc(f) &\geq&
\inf_{f \in \F_M} \Hcr(\r_f)
= \inf_{\r \in \F_M^r} \Hcr(\r) \nonumber \\
&=&
\Hcr(\r_0) = \Hc (f_0) \geq \inf_{f \in \F_M} \Hc(f). \label{infeq}
\eea
Now take any minimizer $g_0 \in \F_M$ of $\Hc$. 
Then by (\ref{infeq}) and part (a),
\[
\inf_{\r \in \F_M^r} \Hcr(\r) = 
\inf_{f \in \F_M} \Hc(f) = \Hc (g_0) = \Hcr (\r_{g_0}),
\]
that is, $\r_{g_0}\in\F_M^r$ minimizes $\Hcr$,
and the proof of part (c) is complete. \prfe

\noindent
{\bf Remark.} If we define an intermediate functional
\[
\P(\r):=\inf_{f\in \F_\r} 
\int\!\!\int\left(\frac{1}{2}|v|^2 f(x,v) + 
Q(f(x,v))\right)\,dv\,dx
\]
with $\F_\r$ as defined in (\ref{frhodef}) then (\ref{intermest})
shows that
\[
\Hc(f) \geq \P(\r_f) + \epot (\r_f) \geq 
\int \Phi(\r_f(x))\, dx + \epot (\r_f) = \Hcr(\r_f)
\]
with equality for minimizers. Note that
$\P(\r)$ is obtained by minimizing the positive contribution
to $\Hc$, which also happens to be the part
depending on phase space densities $f$ directly, over all $f$'s
which generate a given spatial density $\r$. Then in a second
step one minimizes for each point $x$ over all functions
$g=g(v)$ which have as integral the value $\r(x)$.

These constructions are borrowed from \cite{Wo} where they
appear for the special case $Q(f)=f^{1+1/k}$. In \cite{Wo}
the resulting functional of $\r$ is investigated
under the assumption of spherical symmetry by rewriting it
as a functional 
of $m_\r (r) := 4 \pi \int_0^r s^2 \r(s)\, ds$
where $r:=\n{x}$. While minimizers of the present
variational problems are spherically symmetric a posteriori,
the a priori restriction to spherical symmetry implies that
any stability result derived from their minimizing property 
is restricted to spherically symmetric perturbations,
which is undesirable. Moreover, in the last section we will
comment on some extensions of the present techniques to situations
where the minimizers are not spherically symmetric.

\section{Concentration-compactness principle and existence of 
minimizers}
\setcounter{equation}{0}

In this section we prove a concentration-compactness principle
that will yield a solution to the following variational problem:
Minimize the functional
\[
\Hcr (\r) := \int \Phi(\r(x))\, dx + \epot(\r)
\]
over the set
\be \label{spacerdef}
\F_M^r := \Bigl\{ \r \in L^1_+ (\R^3) 
\mid
\int \Phi(\r) < \infty, \ \int \r = M \Bigr\}
\ee
for $M>0$ given 
and $\Phi$ satisfying the following

\smallskip
\noindent {\bf Assumptions on $\Phi$}: 
$\Phi \in C^1 ([0,\infty[)$, $\Phi(0)=0=\Phi'(0)$, and
\begin{itemize}
\item[$(\Phi 1)$]
$\Phi$ is strictly convex. 
\item[$(\Phi 2)$]
$\Phi (\r) \geq C \r^{1+1/n},\ \r \geq 0$ large,  with $0 < n < 3$,
\item[$(\Phi 3)$]
$\Phi (\r) \leq C \r^{1+1/n'},\ \r \geq 0$ small,  with $0 < n' < 3$.
\end{itemize} 
Note that Lemma~\ref{phiprop} tells us that the function $\Phi$ which 
we constructed from a given $Q$ in Section~2 has these properties,
provided $Q$ satisfies the growth conditions corresponding to
$(\Phi 2)$ and $(\Phi 3)$. The aim of this section is to prove the following
result:

\begin{theorem} \label{ccp}
The functional $\Hcr$ is bounded from below on $\F_M^r$.
Let $(\r_i)\subset \F_M^r$ be a minimizing sequence 
of $\Hcr$. Then there exists a sequence of shift vectors 
$(a_i)\subset \R^3$ and a subsequence, again denoted
by $(\r_i)$, such that for any $\epsilon >0$ there exists $R>0$
with
\[
\int_{a_i + B_R} \r_i (x)\, dx \geq M - \epsilon,\ i \in \N,
\]
\[
T \r_i := \r_i(\cdot + a_i) \rightharpoonup \r_0
\ \mbox{weakly in}\ L^{1+1/n}(\R^3),\ i \to \infty,
\]
and
\[
\int_{B_R} \r_0 \geq M - \epsilon .
\]
Finally,
\[
\nabla U_{T\r_i} \to \nabla U_0\ 
\mbox{strongly in}\ L^2(\R^3),\ i \to \infty,
\] 
and $\r_0 \in \F_M^r$ is a minimizer of $\Hcr$.
\end{theorem}

Here and in the following we denote for $0<R<S\leq \infty$,
\beas
B_R 
&:=& \{x \in \R^3 | \n{x} \leq R\},\\ 
B_{R,S} 
&:=& 
\{ x \in \R^3 | R \leq \n{x} < S \}.
\eeas
We split our argument into a series of lemmas.
The first thing to note is that $\Hcr$ is bounded from below on 
$\F_M^r$:

\begin{lemma} \label{lowbound}
Under the above assumptions on $\Phi$,
\[
\Hcr (\r) \geq \int \Phi(\r)\, dx - C -
C \left(\int \Phi(\r)\, dx\right)^{n/3},
\ \r \in \F_M^r,
\]
in particular,
\[
h_M^r := \inf_{\F_M^r} \Hcr > -\infty.
\]
\end{lemma}

\proof
By the extended Young's inequality, interpolation, 
and assumption ($\Phi 2$),
\beas
-\epot (\r) 
&\leq& 
C \nn{\r}_{6/5}^2 \leq C 
\nn{\r}_1^{(5-n)/3} \nn{\r}_{1+1/n}^{(n+1)/3} \\
&\leq&
C+ C \left(\int \Phi(\r)\, dx\right)^{n/3},\ \r \in \F_M^r.
\eeas
Since $n<3$, $\Hcr$ is bounded from below on $\F_M^r$. 
\prfe

\begin{cor} \label{bdminseq}
Any minimizing sequence for $\Hcr$ in $\F_M^r$ is bounded
in $L^{1+1/n}(\R^3)$ and therefore has a subsequence which
converges weakly in $L^{1+1/n}(\R^3)$.
\end{cor}

\proof
By Lemma~\ref{lowbound}, $\int \Phi(\r)$ is bounded along any
minimizing sequence. The assertion follows by $(\Phi 2)$
and the fact that $\int \r = M$ for $\r \in \F_M^r$.
\prfe

Note that the estimates above show that the
definition (\ref{spacerdef}) coincides with our earlier definition
for the set $\F_M^r$.
We also see that the assumption $(\Phi 2)$ is quite natural. 
Next we prove a splitting estimate which will show that along a minimizing
sequence the mass cannot vanish:

\begin{lemma} \label{split1}
Let $\r \in \F_M^r $. Then
\[
\sup_{a\in\R^3} \int_{a+B_R} \r (x)\, dx \geq 
\frac{1}{R M} 
\left( - 2 \epot(\r) - \frac{M^2}{R} - 
\frac{C \nn{\r}_{1+1/n}}{R^{(5-n)/(n+1)}} \right),\ R>1.
\]
\end{lemma}

\proof
We split the potential energy as follows:
\beas
- 2 \epot(\r) 
&=& 
\int\!\!\int_{\n{x-y}\leq 1/R} 
\frac{\r(x)\, \r(y)}{\n{x-y}}\, dx \, dy
+ \int\!\!\int_{1/R < \n{x-y} <  R} \ldots
+ \int\!\!\int_{R \geq \n{x-y}} \ldots \\
&=:&
I_1 + I_2 + I_3.
\eeas
By H\"older's inequality and Young's inequality,
\beas
I_1
&\leq&
\nn{\r}_{1+1/n} \nn{\r \ast ({\bf 1}_{B_{1/R}} 1/\n{\cdot})}_{n+1} 
\leq \nn{\r}_{1+1/n}^2 \nn{{\bf 1}_{B_{1/R}} 1/\n{\cdot}}_{(n+1)/2}\\
&\leq&
C \, \nn{\r}_{1+1/n}^2 R^{-(5-n)/(n+1)};
\eeas
here ${\bf 1}_S$ denotes the indicator function of the 
set $S\subset \R^3$.
The estimates for $I_2$ and $I_3$ are straight forward:
\[
I_2 \leq R \int\!\!\int_{\n{x-y}\leq R} \r(x)\, \r(y)\, dx \, dy
\leq M \, R\, \sup_{a \in \R^3} \int_{a + B_R} \r (x)\, dx,
\]
and
\[
I_3 \leq R^{-1} M^2.
\]
Putting these estimates together yields the assertion.
\prfe

Note that to obtain this estimate we actually split the
Green's function $1/\n{x}$. To exploit this estimate along
minimizing sequences we need to know that $h_M^r< \infty$.
It is here that we need the assumption $(\Phi 3)$:

\begin{lemma} \label{scaling}
\begin{itemize}
\item[{\rm (a)}]
For every $M>0$ we have $h_M^r < 0$.
\item[{\rm (b)}]
For every $0<\bar M \leq M$ we have 
$h_{\bar M}^r \geq (\bar M/M)^{5/3} h_M^r$.
\end{itemize}
\end{lemma}

\proof
For $\r \in \F_M^r$ and $a, b >0$ we define 
$\bar \r(x):= a \r(b x)$.
Then
\beas
\int \bar \r\, dx
&=&
a b^{-3} \int \r \, dx,\\
\epot(\bar \r)
&=&
a^2 b^{-5} \epot (\r),\\
\int \Phi (\bar \r)
&=&
b^{-3}\int \Phi(a \r)\, dx.
\eeas
To prove part (a) we fix a bounded
and compactly supported function $\r \in \F_M^r$ 
and choose $a = b^3$ so that $\bar \r \in \F_M^r$ as well. 
By $(\Phi 3)$ and since $3/n' > 1$,
\[
\Hcr(\bar \r) 
=
b^{-3} \int\Phi(b^3 \r)\, dx + b \, \epot(\r)
\leq
C\, b^{3/n'}   + b \, \epot(\r)
< 0,\ b \to 0,
\]
and part (a) is established. 
As to part (b), we take $a=1$ and 
$b=(M/\bar M)^{1/3} \geq 1$. For $\r \in \F_M^r$ and  
$\bar \r \in \F_{\bar M}^r$ rescaled with these parameters we find that
\beas
\Hcr (\bar \r)
&=&
b^{-3} \int \Phi(\r)\, dx + b^{-5} \epot (\r)\\ 
&\geq&
b^{-5} \left(\int \Phi(\r)\, dx + \epot (\r) \right) =
\left(\frac{\bar M}{M}\right)^{5/3} \Hcr (\r).
\eeas
Since for the present choice of $a$ and $b$
the map $\r \mapsto \bar \r$ is one-to-one and onto between
$\F_M^r$ and $\F_{\bar M}^r$ 
this estimate proves part (b). 
\prfe 

\begin{cor} \label{novanishing}
Let $(\r_i)\subset \F_M^r$ be a minimizing sequence
of $\Hcr$. Then there exist $\delta_0>0$, $R_0>0$, $i_0 \in \N$,
and a sequence of shift vectors $(a_i)\subset \R^3$
such that 
\[
\int_{a_i+B_R} \r_i (x)\, dx \geq \delta_0,
\ i \geq i_0,\ R\geq R_0.
\]
\end{cor}

\proof
By Corollary~\ref{bdminseq},
$(\nn{\r_i}_{1+1/n})$ is bounded.
By Lemma~\ref{scaling} (a) we have
\[
\epot (\r_i) \leq \Hcr (\r_i) \leq 
\frac{1}{2} h_M^r < 0,\ i \geq i_0,
\]
for a suitable $i_0 \in \N$.
Thus by Lemma~\ref{split1} there exist $\delta_0 >0$, $R_0>0$,
and a sequence of shift vectors $(a_i) \subset \R^3$ as required.
\prfe

Finally, we will also need to exploit the well known compactness properties
of the solution operator of the Poisson equation:

\begin{lemma} \label{potcomp}
Let $(\rho_i) \subset L^{1+1/n} (\R^3)$ be bounded and
\[
\rho_i \rightharpoonup \rho_0 \ \mbox{weakly in}\ L^{1+1/n} (\R^3) .
\]
\begin{itemize}
\item[{\rm (a)}]
For any $R>0$,  
\[
\nabla U_{{\bf 1}_{B_R} \rho_i} \to \nabla U_{{\bf 1}_{B_R} \rho_0} 
\ \mbox{strongly in}\ L^2 (\R^3).
\]
\item[{\rm (b)}]
If in addition $(\r_i)$ is bounded in $L^1(\R^3)$, $\r_0 \in \L^1(\R^3)$, and
for any $\epsilon >0$ there exists $R>0$ and $i_0 \in \N$
such that
\[
\int_{\n{x} \geq R} |\r_i(x)| \, dx < \epsilon,\ i \geq i_0
\]
then
\[
\nabla U_{\rho_i} \to \nabla U_{\rho_0} 
\ \mbox{strongly in}\ L^2 (\R^3).
\]
\end{itemize}
\end{lemma}

\noindent
{\bf Proof}. As to part (a), take any $R'>R$. 
Since $1+1/n > 4/3 > 6/5$,
the mapping
\[
L^{1+1/n} (\R^3) \ni \rho \mapsto {\bf 1}_{B_{R'}} 
\nabla U_{{\bf 1}_{B_R}\rho} 
\in L^2(B_{R'})
\]
is compact. Thus the asserted strong convergence holds on $B_{R'}$.
On the other hand,  
\[
\int_{|x| \geq R'} |\nabla U_{{\bf 1}_{B_R} \rho_i}|^2 dx
\leq \frac{C}{R'-R} \|{\bf 1}_{B_R} \rho_i\|_1^2 \leq \frac{C}{R'-R},
\ i \in \N \cup \{0\},
\]
which is arbitrarily small for $R'$ large. As to part (b), 
we have for any $R>0$,
\[
\nn{\nabla U_{\r_i} - \nabla U_{\r_0}}_2
\leq
\nn{\nabla U_{{\bf 1}_{B_R}\r_i} - \nabla U_{{\bf 1}_{B_R}\r_0}}_2 
+
\nn{\nabla U_{{\bf 1}_{B_{R,\infty}}\r_i} - 
\nabla U_{{\bf 1}_{B_{R,\infty}}\r_0}}_2.
\]
Using the extended Young's inequality, interpolation, and the boundedness
of the sequence in $L^{1+1/n}(\R^3)$ we find that
\beas
\nn{\nabla U_{{\bf 1}_{B_{R,\infty}}\r_i} - 
\nabla U_{{\bf 1}_{B_{R,\infty}}\r_0}}_2
&\leq&
C \left(\nn{{\bf 1}_{B_{R,\infty}}\r_i}_{6/5}
+ \nn{{\bf 1}_{B_{R,\infty}}\r_0}_{6/5} \right) \\
&\leq&
C \left(\nn{{\bf 1}_{B_{R,\infty}}\r_i}_1^{(5-n)/6}
+ \nn{{\bf 1}_{B_{R,\infty}}\r_0}_1^{(5-n)/6} \right).
\eeas
Given $\epsilon >0$ we now choose $R>0$ and $i_0\in \N$
such that this is less than $\epsilon >0$ for $i\geq i_0$,
and recalling (a) completes the proof.
\prfe

We are now ready to prove the main result of this section:

\noindent
{\bf Proof of Theorem~\ref{ccp}.}\\
We split $\r \in \F_M^r$ into three different parts:
\[
\r = {\bf 1}_{B_{R_1}} \r + {\bf 1}_{B_{R_1,R_2}} \r + 
{\bf 1}_{B_{R_2,\infty}} \r
=: \r_1 + \r_2 + \r_3;
\]
the parameters $R_1 < R_2$ of the split are yet to be determined. With
\[
I_{lm}:= \int\!\!\int\frac{\r_l (x)\, \r_m (y)}{\n{x-y}},\
l, m = 1,2,3,
\]
we have
\[
\Hcr(\r) = \Hcr(\r_1) + \Hcr(\r_2) + \Hcr(\r_3)
- I_{12} - I_{13} - I_{23} .
\]
If we choose $R_2 > 2 R_1$
then
\[
I_{13} \leq \frac{C}{R_2} .
\]
Next, we use the Cauchy-Schwarz inequality, the extended Young's
inequality, and interpolation to get
\beas
I_{12} + I_{23}
&=&
\frac{1}{4\pi}
\left|\int \nabla(U_1+U_3)\cdot \nabla U_2 dx \right| 
\leq
C \nn{\r_1 + \r_3}_{6/5} \nn{\nabla U_2}_2 \\
&\leq&
C \nn{\r}_{1+1/n}^{(n+1)/6} \, \nn{\nabla U_2}_2.
\eeas   
Using the estimates above and Lemma~\ref{scaling} (b)
we find with $M_l=\int \r_l,\ l=1,2,3$,
\bea \label{split2}
h_M^r - \Hcr(\r)
&\leq&
\left(1-\left(\frac{M_1}{M}\right)^{5/3}
- \left(\frac{M_2}{M}\right)^{5/3}
- \left(\frac{M_3}{M}\right)^{5/3}\right) \, h_M^r 
\nonumber \\
&&
{} + C\, \left(R_2^{-1} + 
\nn{\r}_{1+1/n}^{(n+1)/6} \, \nn{\nabla U_2}_2\right) 
\nonumber \\ 
&\leq&
\frac{C}{M^2}\left(M_1 M_2 + M_1 M_3 + M_2 M_3\right)\, h_M^r
\nonumber \\
&&
{} + C \left(R_2^{-1} + 
\nn{\r}_{1+1/n}^{(n+1)/6} \, \nn{\nabla U_2}_2\right)
\nonumber \\
&\leq&
C h_M^r \, M_1\, M_3 + 
C\, \left(R_2^{-1} + \nn{\r}_{1+1/n}^{(n+1)/6} \, 
\nn{\nabla U_2}_2 \right);
\eea
observe that by Lemma~\ref{scaling} (a) $h_M^r < 0$
and that constants denoted by $C$ are positive and depend
on $M$ and $\Phi$, but not on $R_1$ or $R_2$. 
We want to use (\ref{split2}) to show that up to a 
subsequence and a shift $M_{3}$ becomes small along 
any minimizing sequence  
for $i$ large provided the splitting parameters are suitably chosen. 

The sequence $T\r_i:=\r_i(\cdot + a_i),\ i \in \N$, is minimizing
and bounded in $L^{1+1/n}(\R^3)$ so there exists a
subsequence, denoted by $(T\r_i)$ again, such that
$T\r_i \rightharpoonup \r_0$ weakly in $L^{1+1/n}(\R^3)$, 
cf.\ Corollary~\ref{bdminseq}.
Now choose $R_0 < R_1$ so that by Corollary~\ref{novanishing},
$M_{i,1} \geq \delta_0$ for $i$ large.  
By (\ref{split2}),
\be \label{split3}
-C\, h_M^r \delta_0 M_{i,3} 
\leq
\frac{C}{R_2} + C\, \nn{\nabla U_{0,2}}_2 
+ C \nn{\nabla U_{i,2} - \nabla U_{0,2}}_2 
+ \Hcr(T\r_i) - h_M^r
\ee
where $U_{i,l}$ is the potential induced by $\r_{i,l}$
which in turn has mass $M_{i,l}$, $i\in \N \cup\{0\}$,
and the index $l=1,2,3$ refers to the splitting. 
Given any
$\epsilon >0$ we increase $R_1 >R_0$ such that
the second term on the right hand side of (\ref{split3})
is small, say less than $\epsilon/4$. Next choose
$R_2 > 2 R_1$ such that the first term is small.
Now that $R_1$ and $R_2$ are fixed, the third term in (\ref{split3}) 
converges to zero by Lemma~\ref{potcomp} (a).
Since $(T\r_i)$ is minimizing the remainder in (\ref{split3})
follows suit.
Therefore, for $i$ sufficiently large,
\be \label{concentr}
\int_{a_i + B_{R_2}} T\r_i = M - M_{i,3}
\geq M - (-C\, h_M^r \delta_0)^{-1} \epsilon .
\ee
Clearly, $\r_0 \geq 0$ a.~e..
By weak convergence we have that for any $\epsilon >0$
there exists $R>0$ such that
\[
M \geq \int_{B_R} \r_0\, dx \geq M-\epsilon
\]
which in particular implies that $\r_0 \in L^1(\R^3)$
with $\int \r_0 dx = M$. 
The functional $\r \mapsto \int \Phi (\r)\, dx$ is convex,
so by Mazur's Lemma and Fatou's Lemma
\[
\int \Phi (\r_0)\, dx \leq 
\limsup_{i\to \infty} \int \Phi (T \r_i)\, dx .
\]
The strong convergence
of the gravitational fields now follows by Lemma~\ref{potcomp} (b),
and in particular,
\[
\Hcr (\r_0) \leq \limsup_{i\to \infty} \Hcr(\r_i) = h_M^r
\]
so that $\r_0$ is a minimizer of $\Hcr$.  
\prfe

\section{Applications, symmetries, extensions}
\setcounter{equation}{0}

Although the main purpose of the present paper is to get
a more general understanding of the techniques developed
in \cite{G1,G2,GR1,GR2,GR3,R1,R2} we want to at least
indicate some possible applications
of these techniques. First we should mention that \cite{GR3}
differs from the other papers in so far as there the Casimir
functional is used as part of the constraint under which then the total
energy is minimized. This made it possible to relax the growth conditions
on $Q$---$0<k\leq7/2$ is covered in \cite{GR3}---,but since in the
reduction process we turn $\C(f) + \ekin(f)$ into a new functional
of $\r$, \cite{GR3} seems to be outside the present framework.  

We start with the observation, already noted in
Theorem~\ref{reduce}, that if $\r_0 \in \F_M^r$ is a minimizer
of $\Hcr$ with induced potential $U_0$ then
\[
\r_0 = (\Phi')^{-1}_+(E_0 - U_0) := \left\{ 
\begin{array}{ccl}
(\Phi')^{-1}(E_0 - U_0)&,& U_0 < E_0 \\
0 &,& U_0 \geq E_0, 
\end{array}
\right.
\]
and thus
\be \label{genef}
\lap U_0 = 4 \pi (\Phi')^{-1}_+(E_0 - U_0)
\ee
on $\R^3$. 
The corresponding minimizer of $\Hc$, 
\[
f_0 =
\left\{ 
\begin{array}{ccl}
(Q')^{-1}(E_0 - E) &,& E < E_0 \\
0 &,& E \geq E_0, 
\end{array}
\right.
\]          
is a steady state of the Vlasov-Poisson system, since
$E = E(x,v) =\frac{1}{2} |v|^2 + U_0(x)$ is a conserved
quantity for the characteristics of the Vlasov equation with
potential $U_0$ induced by $\r_0 = \r_{f_0}$. Steady states obtained in this
manner have finite mass $M$, which is a necessary property
for physically relevant steady states. 
We remark that the ansatz $f_0=\phi(E_0-E)$
reduces the stationary Vlasov-Poisson system to the semilinear
Poisson equation
\[
\lap U_0 =
4 \pi \int \phi\left(E_0-\frac{1}{2} |v|^2 - U_0 \right)\, dv
\]
which is exactly (\ref{genef}), provided $Q$ can be chosen such that
$(Q')^{-1} = \phi$ on $\R_+$ and $\phi =0$ on $\R_-$. 
As far as the existence of steady states is concerned
our ``reduced'' approach allows us to cover 
$f_0=(E_0-E)_+^k$ with $-1<k<3/2$ which leads to (\ref{genef})
with right hand side $C\, (E_0-U_0)_+^n$ with $n=k+3/2$ in the
permissible range  $]0,3[$; note that the lower bound $k>-1$ is necessary
to make the $v$-integral above converge. With the direct
approach working with $\Hc$ we were restricted to $0< k< 3/2$.

The main feature of steady states obtained as minimizers in this manner
is that their nonlinear stability. Since this is the main point
in the investigations cited above, we do not go into this here.
Instead, we briefly look a the role of symmetries in our 
problem. First we note that for any $\r_0\in \F_M^r$
its spherically symmetric decreasing rearrangement, denoted by $\r_0^\ast$,
also lies in $\F_M^r$ and satisfies 
\[
\int \Phi(\r_0) = \int \Phi(\r_0^\ast),\
\epot (\r_0) \geq \epot(\r_0^\ast)
\]
with equality if and only if $\r_0 = \r_0^\ast (\cdot - x^\ast)$
for some $x^\ast \in \R^3$, cf.\ \cite[Thms.\ 3.7, 3.9]{LL}. 
In particular, any minimizer of $\Hcr$ must be spherically symmetric
with respect to some point in $\R^3$. If we are only interested
in solving (\ref{genef}) or the stationary
Vlasov-Poisson system we therefore loose nothing if we restrict ourselves to
the set of spherically symmetric functions in $\F_M^r$.
The crucial part of the concentration-compactness argument
simplifies considerably under this restriction:

\begin{lemma} \label{ccpsym}
Define 
\[
R_0=-\frac{3}{5}\frac{M^2}{h_M^r} > 0.
\]
Let  $\r \in \F_M^r$ be spherically symmetric, $R>0$,
and
\[
m := \int_{\{\n{x} \geq R\}} \r.
\]
Then the following estimate holds:
\[
\Hcr(\r) \geq h_M^r + \left[\frac{1}{R_0} - \frac{1}{R}\right]
\, (M-m) \, m.
\]
If $R>R_0$ then for any
spherically symmetric minimizing sequence $(\r_i) \subset \F_M^r$ of $\Hcr$,
\[
\lim_{i \to \infty} \int_{\n{x} \geq R} \r_i = 0.
\]
\end{lemma}

\proof
Clearly,
\[
\Hcr (\r) = \Hcr (\r_1) + \Hcr (\r_2) - 
\int \frac{\r_1(x)\, \r_2(y)}{\n{x-y}}dx\, dy,
\]
where $\r_1 = {\bf 1}_{B_R} \r$, $\r_2 = \r - \r_1$.
Due to spherical symmetry,
\beas
\int \frac{\r_1(x)\, \r_2(y)}{\n{x-y}}dx\, dy
&=&
\frac{1}{4 \pi} \int \nabla U_{\r_1} \cdot \nabla U_{\r_2} dx\\
&=&
\int_0^\infty \frac{4 \pi}{r^2} \int_0^r \r_1(s)\, s^2 ds\, 
\frac{4 \pi}{r^2} \int_0^r \r_2(s)\, s^2 ds\,r^2 dr\\
&=&
\int_R^\infty \cdots dr \leq \frac{(M-m)\, m}{R}.
\eeas
Thus by Lemma~\ref{scaling},
\beas
\Hcr (\r)
&\geq&
h_{M-m}^r + h_{m}^r - \frac{(M-m)\, m}{R}\\
&\geq&
\left[ \left(\frac{M-m}{M}\right)^{5/3} + 
\left(\frac{m}{M}\right)^{5/3} \right]
\, h_M^r - \frac{(M-m)\, m}{R}\\
&\geq&
\left[1-\frac{5}{3} \frac{M-m}{M} \frac{m}{M} \right]
\, h_M^r - \frac{(M-m)\, m}{R}
\eeas
which is the first assertion of the lemma;
note that the scaling transformations in the proof of
Lemma~\ref{scaling} preserve spherical symmetry. 
Now take $R>R_0$ and assume that
the second assertion were false so that up to a subsequence,
\[
\lim_{i \to \infty} \int_{\n{x} \geq R} \r_i = m > 0.
\]
Choose $R_i > R$ such that
\[
m_i:= \int_{\n{x} \geq R_i} \r_i = \frac{1}{2} \int_{\n{x} \geq R} \r_i.
\]
By the already established splitting estimate,
\[
\Hcr(\r_i) 
\geq
h_M^r + \left[\frac{1}{R_0} - \frac{1}{R_i}\right]
\, (M-m_i) \, m_i
\geq
h_M^r + \left[\frac{1}{R_0} - \frac{1}{R}\right]
\, (M-m_i) \, m_i,
\]
and with $i \to \infty$,
\[
h_M^r \geq h_M^r + \left[\frac{1}{R_0} - \frac{1}{R}\right]
\, (M-m/2) \, m/2 > h_M^r,
\]
a contradiction.
\prfe

The lemma above now replaces Lemma~\ref{split1}, Corollary~\ref{novanishing},
and the proof of (\ref{concentr}) which relied on the fairly
lengthy argument via (\ref{split2}) and (\ref{split3}).
In addition, we get a somewhat sharper result on the minimizer:
\[
\supp \r_0 \subset B_{R_0}.
\]
That spherical symmetry helps with compactness issues was already
noted in \cite{Str}.
The a-priori restriction to the spherically symmetric case
is undesirable in view of resulting stability assertions: These would
then be restricted to spherically symmetric perturbations.
Moreover, the symmetry simplification cannot be used if one
does not a priori know that the minimizers will be spherically 
symmetric. One example for this situation is the construction
of steady states with axial symmetry, say with respect to the $x_3$-axis,
by making $Q$ in addition depend explicitly on $x_1 v_2 - x_2 v_1$,
the angular momentum with respect to the axis of symmetry.
Exactly the same reduction procedure as before now gives
a function $\Phi$ that depends in addition on $r = \sqrt{x_1^2 + x_2^2}$,
and minimizers will not be spherically symmetric. 
An investigation of axially symmetric steady states and their stability
will be the content of \cite{GR4}. Another situation where
the minimizers will in general not be spherically symmetric arises
if one includes in the Vlasov-Poisson system an exterior gravitational
field, say $U_e = U_{\r_e}$ with some fixed 
$\r_e \in L^1_+ \cap L^{1+1/n} (\R^3)$.
It is quite easy to check that all the analysis carried
out in this paper extends to this case; only the potential energy needs
to be modified accordingly:
\beas
\epot (\r) 
&=& 
- \frac{1}{2} \int\!\!\int \frac{\r(x)\, \r(y)}{\n{x-y}} dx\,dy
-  \int\!\!\int \frac{\r(x)\, \r_e (y)}{\n{x-y}} dx\,dy \\
&=&
\frac{1}{2} \int \r(x) U_{\r}(x)\, dx + 
\int \r(x) U_e (x)\, dx,
\eeas
and if $\r_e$ is not spherically symmetric then neither are
possible minimizers.
If one wishes to study the minimization of $\Hcr$
with $\Phi(\r)$ generalized to $\Phi(x,\r)$ the crucial step
in the analysis which restricts the possible dependence on $x$
is the scaling in Lemma~\ref{scaling}.
\bigskip

\noindent
{\bf Acknowledgment.}
This paper originates from my collaboration with Y.~Guo, Brown University,
whom I would like to thank for many stimulating discussions.
The research was supported by the Erwin Schr\"odinger International
Institute for Mathematical Physics in Vienna.

\end{document}